# RECENT ASTROPHYSICS RESULTS FROM ORELA AND POSSIBLE FUTURE EXPERIMENTS AT ORELA AND SNS

P. E. KOEHLER

*Physics Division, Oak Ridge National Laboratory, Oak Ridge, TN  37831*
*E-mail: koehlerpe@ornl.gov*

I present some recent results from experiments at the Oak Ridge Electron Linear Accelerator (ORELA) and discuss their impact in nuclear astrophysics. I then describe some possible future nuclear astrophysics experiments at ORELA and at the Spallation Neutron Source (SNS) being built in Oak Ridge.  The SNS and ORELA are complementary, world-class facilities and both will be needed for important future experiments in nuclear astrophysics.

## 1    Introduction

The ORELA facility [1] has a long and distinguished history of experiments in nuclear astrophysics. Most of the neutron capture reaction rates used in nuclear astrophysics calculations were determined in experiments at ORELA by Dick Macklin and collaborators. More recently, an improved apparatus [2] has made it possible to measure these rates much more accurately and to lower energies than before. These new data are needed to make use of new high precision isotopic anomaly data from meteorites [3] and to test the latest stellar models [4]. Also, it was recently realized [5] that $(n,\alpha)$ experiments at ORELA could provide perhaps the best constraints on the many $(\gamma,\alpha)$ rates needed for explosive nucleosynthesis calculations. In addition, ORELA would be an excellent facility for several other types of important nuclear astrophysics experiments, such as inelastic neutron scattering, neutron capture on long-lived radioactive samples, or total cross section measurements on shorter lived radioactive samples. Most of these topics are discussed in an ORELA "White Paper" that is available at http://www.phy.ornl.gov/astrophysics/nuc/neutrons/whitepaper.pdf.

The intense neutron flux at the SNS [6] should allow measurements on much smaller samples than is possible at ORELA. A recent study [7] indicates that the flux at the SNS should be over 10000 times higher that at the ORELA "benchmark" facility where most previous neutron capture measurements for nuclear astrophysics have been made, and over 40 times higher than at the new DANCE instrument (see John Ullmann's paper in these proceedings) at the Los Alamos Neutron Science Center (LANSCE) [8]. Therefore, the SNS should make possible measurements on widest range of radioactive and very small stable samples of interest to nuclear astrophysics.

## 2  Recent ORELA Results and the Need for New Measurements

If you are not familiar with this field, I urge you to read the excellent overview paper by Franz Käppeler at the beginning of these proceedings to acquaint yourself with the basic concepts and jargon.

### 2.1  The cool, new s-process models

The latest, most realistic, and most successful models [4] of the *s* process indicate that roughly half of the abundances of nuclides heavier than A≈100 were made in low-mass Asymptotic Giant Branch (AGB) stars. A major difference between these and previous models is that most of the neutron exposure driving the nucleosynthesis occurs at much lower temperatures ($kT$=6-8 keV) than previously thought ($kT$=30 keV). This could be a problem because most of the old measurements, and even some of the new high-precision measurements, were not made to low enough energies to obtain the reaction rates at these new lower temperatures without resorting to extrapolations. At ORELA, we can routinely measure the neutron capture cross sections across the entire range of energies needed. In all [9,10] but one [11] of the cases studied so far, new ORELA measurements indicate that extrapolations from previous data to obtain reaction rates at the low temperatures needed by new stellar models are in error by two to three times the estimated uncertainties. Therefore, extrapolated rates are not sufficiently accurate for meaningful tests of new stellar models. More low-energy measurements are needed, especially for the *s*-only isotopes that serve as the most important calibration points for the models.

### 2.2  Cracks in the Classical s-process Model

The so-called classical model of the *s* process has been used for many years because, by making some simplifying assumptions (constant temperature, neutron density, and matter density), it is possible to find an analytical solution to the large network of time-dependent, coupled differential equations describing the reaction flow during the *s* process. As a result, the classical model has been very useful for ascertaining the mean conditions of the *s*-process environment. Although stellar models indicate that the classical model assumptions are too severe, it has been amazingly successful in reproducing the observed *s*-process abundances.

   The competition between neutron capture and beta decay at several relatively long-lived radioactive nuclides along the *s*-process path can yield a very direct handle with which to estimate the average neutron density, temperature and matter density in the stellar plasma during the *s* process. Because we know assumptions of the classical *s*-process model are too simplistic for real stars, if ($n,\gamma$) cross-section data of sufficient accuracy exist, classical analyses of different branchings should eventually yield inconsistent results.

Thanks to precise new data from ORELA, cracks in classical model are beginning to show. Previous classical analyses of branchings in the *s*-process path had led to a temperature of $kT = 29\pm5$ keV. In contrast, recent precise $^{134,136}$Ba$(n,\gamma)$ reaction rate measurements from ORELA [8] were used in a classical analysis of a different branching to deduce a mean *s*-process temperature of $kT = 15\pm5$ keV. This was the first time that clearly inconsistent temperatures were obtained from different *s*-process branchings. Even more recent ORELA measurements on isotopes of Pt [12] have yielded a second example of an inconsistency, this time in the neutron density. Because stellar models are complicated, more precision measurements near other branching points (e.g., $^{85}$Kr, $^{95}$Zr, $^{151}$Sm, $^{152}$Eu, $^{153}$Gd, $^{163}$Ho, $^{170,171}$Tm,..) will be needed to understand the crucial ingredients in the new stellar models.

Additional cracks in the classical *s*-process model appeared when new ORELA data [12,13] demonstrated for the first time that the classical model of the *s* process fails to predict the correct abundances of the *s*-only isotopes $^{142}$Nd and $^{192}$Pt.

## 2.3 Red Giant Stardust

Microscopic grains of silicon carbide and other refractory materials recovered from primitive meteorites represent a new class of observational data with which to constrain astrophysical models. Most of these grains appear to be actual stardust from red giant (AGB) stars inside of which the *s* process had occurred. Trapped within these grains are trace amounts of several intermediate- to heavy-mass elements having isotopic patterns that are very non-solar. Qualitatively, these patterns agree with the expectations of nucleosynthesis from the *s* process – they are relatively enriched in *s*-process isotopes and depleted in isotopes thought to come from the *r* and *p* processes. Also, the precision to which these isotope ratios can be measured is much higher than the precision of measured element-to-element abundances for the solar system. Therefore, these new meteorite data extend both the number and precision of the calibration points for *s*-process models. New high-precision neutron capture data are needed to see if this beautiful, qualitative red-giant stardust model can be made quantitative.

The first precise test of the red-giant stardust model recently was made possible when new $^{142,144}$Nd$(n,\gamma)$ cross sections were measured [13] with good precision at ORELA. Stellar *s*-process model calculations made with previously accepted cross sections were in serious disagreement with the stardust data. The new ORELA measurements, which were made with an improved apparatus and over a wider energy range, showed that the old data were in error. With the new ORELA data, the agreement between the stellar model and the stardust data was excellent.

Subsequent ORELA measurements for isotopes of barium [9,11] have revealed problems in the red giant stardust model. Stardust data for other elements exist (e.g., Sr, Mo, and Dy), but because many of the existing ($n,\gamma$) data are too imprecise or do not cover the entire energy range needed by the models, new measurements are needed to make use of these data to test and improve the red-giant stardust model.

*2.4  Improving Reaction Rates for Explosive Nucleosynthesis Models*

The neutron-deficient isotopes (the so-called *p* isotopes) of intermediate- to heavy-mass elements cannot be made by neutron capture reactions starting from stable "seed" nuclides. It is thought that they were synthesized when seeds built up by a previous *s* process were photo-eroded in an explosive, high-temperature environment during the *p* process. The site of the *p* process is unknown, but the leading candidates appear to be the late stages in the lives of massive stars or supernova explosions.

The largest nuclear physics uncertainties in these models are the rates for ($\gamma,\alpha$) reactions. Determining these rates through direct laboratory measurements is very difficult if not impossible because the cross sections are extremely small due to high Coulomb barriers. Because the level densities are high at the excitation energies and masses of interest, the rates can, in principle, be calculated to sufficient accuracy using the statistical model. However, the $\alpha$+nucleus potential needed for this model is very poorly constrained, so calculated rates are very uncertain. Traditional methods for constraining potentials are problematical because they involve a largely unconstrained extrapolation.

A series of low-energy ($n,\alpha$) measurements, across a wide range of masses, appears to be the best means of constraining the $\alpha$+nucleus potential and thus improving the calculation of these rates. The Q values for these ($n,\alpha$) reactions are such that the relative energies between the $\alpha$ particle and the residual nucleus are in the astrophysically interesting energy range, so no extrapolation is necessary

At ORELA, the first application of this idea was to measure the $^{143}$Nd and $^{147}$Sm($n,\alpha$) [5] cross sections across the range of energies needed for astrophysics applications. Previous measurements of this type were limited to energies below a few keV (which is too small of an energy range to be useful for comparison to statistical models) due to overload problems in the detectors and associated electronics resulting from the $\gamma$ flash at the start of each neutron pulse. In the new experiments, this problem was overcome by employing a compensated ion chamber (CIC) [14] as the detector. The CIC reduced the $\gamma$-flash background to the point where measurements are possible to much higher energies (500 keV in the cases of $^{143}$Nd and $^{147}$Sm).

Results from the ORELA measurements are shown in Fig. 1. The older calculations of Holmes *et al*. [15] are much closer to the data than the newer NON-SMOKER [16] or MOST [17] calculations, which differ from the data by about a factor of 3 in opposite directions. The better agreement of the older model may be due to a fortuitous cancellation of effects. The newer models employ a neutron potential that is known to be more reliable in this mass region. In addition, the authors of the newer models have attempted to reduce the reliance on empirical fine tuning and to take advantage of the latest physics knowledge in an effort to increase the reliability of the models away from the valley of stability. In the case of the $\alpha$ potential, several parameters are needed to account for the mass, energy, and nuclear structure effects. At present, the values of these parameters in the astrophysically relevant range are poorly constrained by experiment.

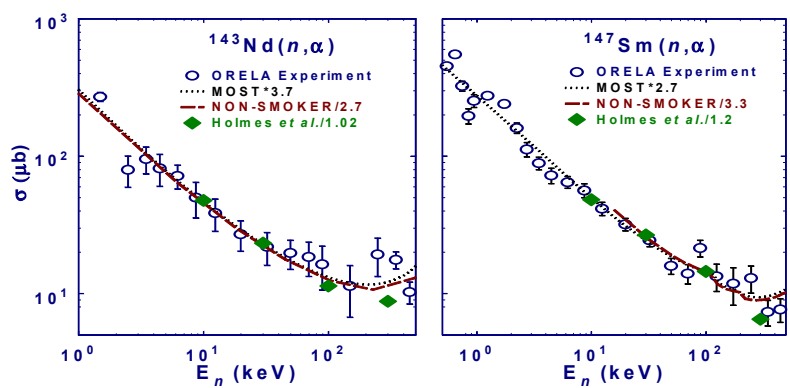

**Figure 1.** Cross sections for the $^{143}$Nd and $^{147}$Sm$(n,\alpha)$ reactions in the unresolved region. Shown are the recent ORELA measurements and the calculations of Holmes *et al*. [15], as well as calculations made with the newer statistical model codes NON-SMOKER [16] and MOST [17]. Note that the theoretical calculations have been normalized by factors given in the legends.

We have studied [5] the sensitivity of calculated $(n,\alpha)$ cross sections to the $\alpha$ potential and level densities employed in the model. We found that differences of about a factor of 30 could be accounted for in the variation of the potential alone. On the other hand, different level density prescriptions changed the cross section by a factor of about 1.5, far smaller than the effect of the $\alpha$ potential. More $(n,\alpha)$ data across as wide a range of masses and energies as possible are needed to constrain the several parameters thought to be necessary to define the global $\alpha$ potential needed for astrophysics applications. Counting rate estimates based on these initial experiments indicate that as many as 30 measurements should be possible across the mass range from S to Hf. However, a new detector that allows higher pressures and voltages, as well as more sample plates will be required for many of these measurements, and some will likely require higher flux than is available at ORELA.

*2.5 Reaction Rates for Thermally Populated Excited States*

The ($n,\gamma$) reaction rates inside the thermal plasma of a star can be significantly different from the rates measured in the laboratory due to reactions involving thermally populated excited states. These stellar enhancement effects cannot be directly measured, but can be determined by measuring neutron inelastic cross sections to the same levels populated in the stellar environment. The enhancement of stellar reaction rates due to this effect can be as large as 30%, and enhancements calculated by various nuclear statistical models [15-17] can differ by substantial amounts. Such large and uncertain effects are particularly troublesome for *s*-only isotopes because they are the main calibration points for *s*-process

The stellar enhancement factor (SEF) for one nuclide along the *s*-process path ($^{187}$Os) was determined [18] through neutron inelastic scattering measurements at ORELA several years ago. This measurement was particularly difficult because the excitation energy of the first excited state in $^{187}$Os is only 9.8 keV; hence, it was not possible to detect the de-excitation $\gamma$ rays directly and it was difficult to resolve the inelastically scattered neutrons from the larger elastic group. However, by using a clever technique that exploited the excellent time-of-flight resolution available at ORELA, it was possible to measure this cross section.

There are four *s*-only isotopes ($^{154}$Gd, $^{160}$Dy, $^{170}$Yd, and $^{176}$Hf) in addition to $^{187}$O calculated to have SEFs greater than 10%, and there are substantial differences between the SEFs calculated by different statistical models. In contrast, current techniques can determine laboratory reaction rates to 1-3% accuracy and isotopic abundances can often be measured with part-per-thousand accuracy. A program of ($n,n'$) measurements for these isotopes is clearly needed to determine enhancements for *s*-only isotopes and to improve statistical models so reliable enhancements can be calculated for other nuclides of interest. Measurements should be easier for these four *s*-only isotopes than for $^{187}$Os because both excitation energies and natural abundances are higher. There are about 25 other nuclides along the *s*-process path calculated to have SEFs larger than 10%. First excited state energies range from 8.4 keV ($^{169}$Tm) to 100 keV ($^{182}$W). For those with the smallest excitation energies it appears as if a technique similar to that used in the $^{187}$Os experiment will be necessary. However, it should be possible to use a flight path about a factor of four shorter (and hence have a higher counting rate or use smaller samples) than in the previous measurement [18] and still obtain resolution sufficient to separate the elastic and inelastic groups.

## 3 The Need for the SNS

Measurements on radioactive samples and on stable samples having very small natural abundances or small cross sections are needed for several reasons. First, measurements on radioactive samples are needed for improving models of the *s* process. For example, neutron capture measurements for radioactive branching points along the *s*-process path could greatly aid in understanding dynamics of the *s* process environment. Almost none of these measurements have been made. Branching points in the *s*-process path for which measurements are needed include $^{85}$Kr, $^{134,135}$Cs, $^{147}$Nd, $^{147,148}$Pm, $^{151}$Sm, $^{152}$Eu, $^{153}$Gd, $^{163}$Dy, $^{163,164}$Ho, $^{169}$Er, $^{170,171}$Tm, $^{176}$Lu, $^{185}$W, and $^{186}$Re.

Second, there are several lighter nuclides whose abundances are modified by (*n*,γ), (*n*,α), and (*n,p*) reactions during the *s* process or during explosive nucleosynthesis. In some cases, these nuclides are of interest to γ-ray astronomy (e.g., $^{22}$Na and $^{26}$Al), to meteoric anomalies (e.g., Si, Cl, Ca, $^{50}$V, and Ti), or to the origin of rare isotopes of lighter nuclides (e.g., $^{36}$Cl, $^{37,39}$Ar). Although measurements exist for many of these cases, the data from different measurements are in serious disagreement, or of poor precision or questionable quality, or cover too limited an energy range for astrophysics applications.

Third, very few measurements of (*n*,γ) reaction rates for nuclides involved in the *p* process have been made, so theoretical rates are used in *p*-process calculations. From measurements made near the valley of stability, it is known that calculated (*n*,γ) rates using global statistical models are accurate to within a factor of two. However, it is not known how reliable the theoretical rates are when extrapolated away from the valley of stability. Measurements on neutron deficient radioactive samples and on the very rare *p*-isotopes themselves are needed to improve reliability of theoretical rates that must still be used for the many cases that cannot be measured. Radioactive nuclides of interest to the *p* process include $^{53}$Mn, $^{55}$Fe, $^{57}$Co, $^{59}$Ni, $^{91,92}$Nb, $^{93}$Mo, $^{97}$Tc, $^{109}$Cd, $^{137}$La, $^{139}$Ce, $^{143,145}$Pm, $^{145,146}$Sm, $^{148,150}$Gd, $^{154,159}$Dy, $^{157}$Tb, $^{172}$Hf, $^{195}$Au, $^{194}$Hg, and $^{202}$Pb.

Fourth, measurements on neutron rich radioactive isotopes are needed to improve models of the *r* process. At the end of the *r* process when reactions freeze out as the temperature and neutron density decline, (*n*,γ) reactions could help smooth out the abundance distribution and improve agreement between astrophysical models and observed *r*-process abundances. The half lives of the involved isotopes are too short for direct measurements, but (*n*,γ) measurements as far off the valley of stability as possible would be very helpful for improving the reliability of theoretical rates. For this reason, reaction rate measurements on the following nuclides are needed: $^{90}$Sr, $^{123}$Sn, $^{126}$Sn, $^{127m}$Te, $^{182}$Hf, $^{210}$Pb, $^{226}$Ra, and $^{227}$Ac.

Minimizing the sample size needed is usually the most important consideration for measurements on very rare stable or on radioactive isotopes. The widest range of measurements of this type should be possible at the SNS because the flux at the SNS is expected to be much larger than at any other facility. High peak flux can also be important in overcoming the background from the decay of the sample, and the SNS also is expected to have the highest peak flux. Recent $^{191,193}$Ir(*n*,γ) [19] and $^{171}$Tm(*n*,γ) [20, Ullmann *et al*., these proceedings] experiments at LANSCE have demonstrated the potential of the high fluxes available at spallation sources for measurements on very small samples. For example, the Ir measurements were made with approximately 1 mg of sample and required a total of only one day of beam time each.

Recently, a comparison [7] was made of various white neutron sources for measurements using radioactive samples. In Table 1, LANSCE [8], CERN-TOF [21], and the SNS [6] are compared to ORELA [1] operating under conditions (8 kW power, 8 ns electron pulse width and 40 m flight path length) typically used in previous ($n,\gamma$) measurements. Neutron capture measurements at the ORELA facility have also been made at flight paths as short as 10 m and the facility has run for extended periods at powers as high as 50 kW; hence, these conditions are also included in Table 1. The LANSCE facility has reliably operated at a power of 64 kW and a flight path length of 20 m has been chosen for the new DANCE instrument [20]. The CERN-TOF facility was originally envisioned with detector stations at 80 and 230 m although currently an original flight path length of 180 m is being instrumented. Under the most optimistic conditions, the power of this facility will be 45 kW. For this comparison, a flight path length of 20 m was chosen for the SNS because this would yield a time-of-flight resolution equivalent to previous nuclear astrophysics measurements at LANSCE.

As can be seen in Table 1, the flux at the SNS is expected to be about 11000 times larger than at the ORELA benchmark facility. Furthermore, the peak flux at the SNS should be larger by a factor of 180000.

**Table 1.** Ratios to the benchmark facility (ORELA at 8 kW power, 8 ns pulse width, and 40 m flight path length) and estimated samples sizes.

| Parameter | Ratio to ORELA Benchmark | | | | | |
|---|---|---|---|---|---|---|
| | ORELA | | LANSCE | CERN-TOF | | SNS |
| Flight Path Length[a] (m) | 10 | 40 | 20 | 230 | 80 | 20 |
| Power | 6.2 | 1 | 8 | 5.6 | 5.6 | 250 |
| Flux at 30 keV | 100 | 1 | 280 | 10 | 85 | 12000 |
| Integral Flux 1 - 300 keV | 100 | 1 | 230 | 9.5 | 75 | 10500 |
| Peak Flux 1 -300 keV | 210 | 1 | 12000 | 520 | 12000 | 180000 |
| 1/(Pulse Width) | 0.083 | 1 | 0.032 | 6.7 | 2.3 | 0.011 |
| Sample Size[b] (mg) | 0.25 | 25 | 0.10 | 2.5 | 0.31 | 0.0022 |

[a]The flight path length is given to define the facility parameters. All other entries except sample sizes are ratios to the benchmark facility.
[b]The samples sizes are scaled to a cross section of 1 b and an atomic number of 150, and assume a detector efficiency of 100%.

Estimates of the sample sizes needed for ($n,\gamma$) measurements are also given in Table 1. The sample sizes have been scaled from previous measurements at the benchmark facility assuming a Maxwellian-averaged cross section at kT=30 keV of 1 b, an atomic number of 150, and a 100% efficient detector. These general numbers can be used to estimate the sample size needed for a particular case by dividing by the calculated 30-keV cross section in b and scaling to the atomic mass. These estimates indicate that it should be possible to measure many of the ($n,\gamma$) reaction rates for radioisotopes of interest to astrophysics at the SNS. In addition, the sample sizes are small enough that the necessary isotopically enriched samples should be affordable for many of the $p$ isotopes. Finally, it should also be possible to measure most of the ($n,\alpha$) cross section of interest to explosive nucleosynthesis.

Although the SNS appears to have much potential for future nuclear astrophysics experiments, there are likely to be a number of challenges to overcome. Producing the radioisotopes and fabricating the radioactive samples will present a significant challenge. In addition, it likely will be difficult to build a detector capable of recovering from the "$\gamma$-flash" at the beginning of each neutron pulse so that measurements can be made to the relatively high energies (few hundred keV) needed. Also, overcoming the backgrounds from the radioactive sample, from neutrons scattered from the sample and its backing, and from other flight paths is expected to be a challenge. Simulations [22] indicate that a detector based on the scintillator $BaF_2$ together with judicious layers of neutron absorbing materials appears to be the best choice.

SNS capabilities are complementary to ORELA and both will be needed to satisfy all the data needs for nuclear astrophysics. For example, in addition to the needed measurements outlined above, ORELA is ideally suited for measuring total cross sections using small amounts of radioisotopes [23]. Even with the large flux available at the SNS, ($n,\gamma$) cross sections for some radioisotopes will remain out of reach for which total cross section measurements should be possible at ORELA. These data could be very useful in constraining statistical model calculations of the ($n,\gamma$) rates.

 Acknowledgements

References

1. R. W. Peele *et al.*, Technical Report No ORNL/TM-8225, Oak Ridge National Laboratory; K. H. Bockhoff *et al*, Nucl. Sci. Eng. **106**, 192 (1990).
2. P. E. Koehler et al., Phys. Rev. C **62**, 055803 (2000).

3. See for example, G. K. Nicolussi *et al*., Phys. Rev. Lett. **81**, 3583 (1998).
4. See for example, C. Arlandini *et al*., Astrophys. J. **525**, 886 (1999).
5. Yu. M. Gledenov *et al*., Phys. Rev. C **62** 042801(R) (2000).
6. D. Olsen *et al*., Technical Report, Oak Ridge National Laboratory Report SNS 100000000PL0001R00, (1999).
7. P. E. Koehler, Nucl. Instr. and Meth. **A460**, 352 (2001).
8. P. W. Lisowski, C. D. Bowman, G. J. Russell, and S. A. Wender, Nucl. Sci. Eng. **106**, 208 (1990).
9. P. E. Koehler *et al*., Phys. Rev. C **54**, 1463 (1996).
10. P. E. Koehler *et al*., Phys. Rev. C **64**, 065802 (2001).
11. P. E. Koehler *et al*., Phys Rev. C **57** R1558 (1998).
12. P. E. Koehler *et al*., to be published in *The Proceedings of the International Conference on Nuclear Data for Science and Technology*, Oct. 7-12, 2001, Tsukuba, Japan.
13. K. H. Guber, R. R. Spencer, P. E. Koehler, and R. R. Winters, Phys. Rev. Lett. **78**, 2704 (1997).
14. P. E. Koehler, J. A. Harvey, and N. W. Hill, Nucl. Instr. and Meth. **A361**, 270 (1995).
15. J. A. Holmes, S. E. Woosley, W. A. Fowler, and B. A. Zimmerman, At. Data Nucl. Data Tables **18**, 305 (1976).
16. T. Rauscher and F.-K. Thielemann, in *Stellar Evolution, Stellar Explosions, and Galactic Chemical Evolution*, Edited by A. Mezzacappa (Institute of Physics, Bristol, 1998), p. 519.
17. S. Goriely, in *Nuclei in the Cosmos*, edited by N. Prantzos and S. Harissopulos (Editions Frontieres, Gif-sur-Yvette, 1998) p. 484.
18. R. L. Macklin, R. R. Winters, N.W. Hill, and J. A. Harvey, Astrophys. J. **274**, 408 (1983).
19. P. E. Koehler and F. Käppeler, in *International Conference on Nuclear Data for Science and Technology*, edited by J. K. Dickens (American Nuclear Society, La Grange Park, 1994) p. 179.
20. J. L. Ullmann *et al*., in *Proceedings of the Fifteenth International Conference on the Application of Accelerators in Research and Industry*, edited by J. L. Guggan and I. L. Morgan (American Institute of Physics, New York, 1999) p. 251.
21. C. Rubbia *et al*., Technical report, European Laboratory for Particle Physics Report CERN/LHC/90-02 (EET), 1998; CERN/SPSC 99-8 SPSC/P 310, 1999.
22. M. Heil, *et al*., Nucl. Instr. Meth. **A459**, 229-246 (2001).
23. R. W. Benjamin *et al*., Nucl. Sci. Eng. **85**, 261 (1983).